\documentclass{PoS}

\title{Tests of hadronic vacuum polarization fits for the muon anomalous magnetic moment}

\ShortTitle{Tests of hadronic vacuum polarization fits for the muon $g-2$}

\author{Maarten Golterman\\
        Department of Physics and Astronomy,
San Francisco State University\\ San Francisco, CA 94132, USA\\
Institut de F\'\i sica d'Altes Energies (IFAE),
Universitat Aut\`onoma de Barcelona\\ E-08193 Bellaterra, Barcelona, Spain\\
        E-mail: \email{maarten@sfsu.edu}}

        \author{Kim Maltman\\
        Department of Mathematics and Statistics,
York University\\  Toronto, ON Canada M3J~1P3
\\
CSSM, University of Adelaide, Adelaide, SA~5005 Australia\\
        E-mail: \email{kmaltman@yorku.ca}}

\author{\speaker{Santiago Peris}
\\
        Department of Physics, Universitat Aut\`onoma de Barcelona\\ E-08193 Bellaterra, Barcelona, Spain\\
        E-mail: \email{peris@ifae.es}}

\abstract{We construct a physically motivated model for the isospin-one
non-strange vacuum polarization function $\Pi(Q^2)$ based on a spectral
function given by vector-channel OPAL data from hadronic $\tau$ decays
for energies below the $\tau$ mass and a successful parametrization, employing
perturbation theory and a model for quark-hadron duality violations, for
higher energies. Using a covariance matrix and $Q^2$ values from a
recent lattice simulation, we then generate fake data for $\Pi(Q^2)$ and
use it to test fitting methods currently employed on the lattice for
extracting the hadronic vacuum polarization contribution to the muon
anomalous magnetic moment. This comparison reveals a systematic error
much larger than the few-percent total error sometimes claimed for such
extractions in the literature. In particular, we find that errors deduced
from fits using a Vector Meson Dominance ansatz are misleading, typically
turning out to be much smaller than the actual discrepancy between the
fit and exact model results. The use of a sequence of Pad\'{e} approximants,
recently advocated in the literature, appears to provide a safer fitting strategy.}

\FullConference{31st International Symposium on Lattice Field Theory - LATTICE 2013\\
		July 29 - August 3, 2013\\
		Mainz, Germany}

\begin{document}

\section{Introduction}

\begin{figure}
\centering
\includegraphics[width=2in]{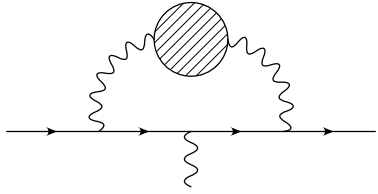}
\caption{Vacuum polarization contribution to $a_{\mu}$. The shaded blob
contains only quarks and gluons, wavy lines are photons and the solid
line is the muon.}
\label{vacpol}
\end{figure}

Recent measurements of the muon anomalous magnetic moment, $a_\mu =(g-2)/2$,
have reached an unprecedented level of precision~\cite{Bennett}. In the
Standard Model (SM), light-quark loop contributions strongly
weighted at low-energy make the hadronic contribution to $a_\mu$
nonperturbative in the strong coupling $\alpha_s$. A numerically
important example is the hadronic vacuum polarization (HVP) contribution
depicted in Fig. \ref{vacpol}. This contribution can be evaluated as an
appropriately weighted integral over the electromagnetic spectral
function, $\rho_{EM}(s)$, and, luckily, since Nature has solved QCD,
$\rho_{EM}$ can be determined experimentally from measured
$\sigma(e^+e^-\rightarrow hadrons)$ data~\cite{exp}.
Adding all other contributions, one finds a SM prediction
for $a_\mu$ more than 3 sigma away from experiment~\cite{Bennett}, opening
the door to beyond-the-SM speculations. Given the stakes, a reevaluation
of this result using completely different techniques is of considerable
interest, and the lattice stands out as the only obvious
systematic nonperturbative theoretical approach available.

On the lattice, the HVP contribution is given by the integral
\begin{equation}\label{one1}
    a_{\mu}^{\rm HVP} = 4\alpha^2\int_0^\infty dQ^2\,f(Q^2)\ \big[\Pi(0)-\Pi(Q^2)\big]\
\end{equation}
with $Q^2$ the Euclidean squared momentum and $f(Q^2)$ a known
kernel~\cite{Blum}. A key feature of Eq. (\ref{one1}) is the very strong
peaking of the integrand near $Q^2\sim m^2_{\mu}/4\sim 0.003\ \mathrm{GeV}^2$, a
scale well below the smallest $Q^2$ reachable on typical current lattices
(e.g., for the data set considered in Ref.~\cite{ABGP}, with
$1/a=3.3554\ \mathrm{GeV}$ and $T=144$,
$Q^2_{min}\sim (2\pi/aT)^2\sim 0.02143\ \mathrm{GeV}^2$).
This feature precludes evaluating Eq.~(\ref{one1}) as a
Riemann sum, making the preliminary step of fitting $\Pi (Q^2)$ to a known
function, which can then be used to evaluate the integral, unavoidable. If
the chosen fit function is not capable of reproducing accurately enough the
true $\Pi(Q^2)$ in the region of the peak of the integrand, this step
introduces a systematic error which contaminates the final result for $a_\mu$.
It is of course very difficult to assess this systematic error without
knowing the true $\Pi(Q^2)$. A good model can help provide a benchmark
for assessing this systematic error. We briefly
summarize the analysis based on such a model, which has recently appeared in
Ref. \cite{GMP}, in what follows.

\section{The Model}

\begin{figure}
\centering
\includegraphics[width=2.5in]{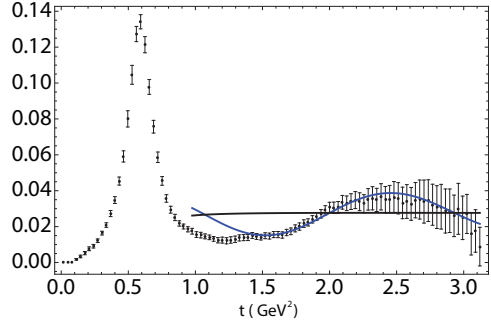}
\caption{The $I=1$ non-strange vector spectral function, from
Ref.~\cite{alphasus}. Experimental data is from OPAL~\cite{OPAL}.
The black solid line is the result of perturbation theory,
$\rho_{\rm pert}(t)$, and the blue solid line the result after
adding our parametrization of duality violations, $\rho_{\rm DV}(t)$,
described in the text.}
\label{spectrum}
\end{figure}

We define the model vacuum polarization function by the once-subtracted
dispersion relation
\begin{equation}\label{two}
  \Pi(Q^2)=-Q^2\int_{4 m_{\pi}^2}^\infty dt\;\frac{\rho(t)}{t(t+Q^2)}\ ,
\end{equation}
with the spectral function $\rho(t)$ determined as follows. In the region
$4m_{\pi}^2\le t\le s_{min}\le m_{\tau}^2$, we use the non-strange $I=1$
vector-channel OPAL spectral data, updated for current branching
fractions, while, for $s_{min}\le t<\infty$, we use the parametrization
\begin{equation}\label{three}
    \rho_{t\ge s_{min}}(t)= \rho_{\rm pert}(t)+\rho_{\rm DV}(t) \quad , \quad
    \rho_{\rm DV}(t) = e^{-\delta-\gamma\, t}\sin(\alpha+\beta t)\ ,
\end{equation}
where $\rho_{\rm pert}$ is the spectral function calculated to five loops
($O\left(\alpha_s^4\right)$) in perturbation theory~\cite{Baikov},
and $\rho_{\rm DV}$ models effects due to quark-hadron duality violations.
These duality violations are due to the presence of resonances in the
spectrum and cannot be captured by the Operator Product Expansion. The
values used for the parameters $s_{min}, \alpha, \beta, \delta, \gamma$
can be found in Ref.~\cite{GMP}. This parametrization of $\rho (t)$ was
developed in Ref.~\cite{DVus}, building on earlier ideas discussed in
Ref.~\cite{Shifman}, and has been extensively studied in
Refs.~\cite{alphasus} in the context of a determination of $\alpha_s$
from hadronic $\tau$ decay data. Figure~\ref{spectrum} shows how well
this parametrization is able to describe the experimental data.

To focus our tests on the potentially problematic, very low $Q^2$ region
of the $a_{\mu}^{\rm HVP}$ integral in (\ref{one1}), we use as benchmark
the $Q^2\le 1\  \mathrm{GeV}^2$ contributions in the model
\begin{equation}\label{four}
\widetilde{a}_{\mu}^{\mathrm{HVP}, Q^2\leq  1}=
4\alpha^2\int_0^{1\mathrm{GeV}^2}
 dQ^2\,f(Q^2)\ \big\{\Pi(0)-\Pi(Q^2)\big\}= 1.204 \times 10^{-7}\ .
\end{equation}
Results using various fit strategies will be compared to this below.
The tilde in $\widetilde{a}_{\mu}^{\rm HVP}$ distinguishes the $I=1$
model result from the true value in nature, $a_{\mu}^{\rm HVP}$.
As can be seen from Eq.~(\ref{two}), the model $\Pi (Q^2)$ is already
subtracted at $Q^2=0$ and hence satisfies $\Pi(0)=0$.

\section{Generation of fake lattice data and fit functions}

We focus on the discrete set $\{Q_i^2\}$ of $Q^2$ values available on a
$L^3\times T=64^3\times 144$ lattice with periodic boundary conditions and
inverse lattice spacing $1/a=3.3554$ GeV~\cite{ABGP}. The smallest
squared momentum is then $(2\pi/a T)^2=0.02143\ \mathrm{GeV}^2$,
significantly larger than the scale $\sim m^2_{\mu}/4= 0.003\ \mathrm{GeV}^2$
at which the strong peak of the integrand in Eq.~(\ref{one1}) occurs (cf.
Fig.~\ref{moral} below). For these $Q_i^2$ we have constructed a multivariate
Gaussian distribution with central values $\Pi(Q_i^2)$ and a covariance
matrix set equal to that of the real lattice data of Ref.~\cite{ABGP}. A
fake data set is then obtained by drawing one random sample from this
distribution. This sample will be the one to be used in all our comparisons
below.

As a family of fit functions, we use the sequence of Pad\'{e}
approximants described in Ref.~\cite{ABGP},
\begin{equation}\label{five}
    \Pi(Q^2)=\Pi(0)-Q^2\left(a_0+\sum_{k=1}^K\frac{a_k}{b_k+Q^2}\right)\ .
\end{equation}
For $a_0=0$ the term in parentheses is a $[K-1,K]$ Pad\'{e}, whereas if $a_0$
is a free parameter we have a $[K,K]$ Pad\'{e}. This sequence is known to
converge to the vacuum polarization (VP) function.

As noted above, the model VP function, $\Pi (Q^2)$, is
already subtracted at $Q^2=0$ and satisfies $\Pi(0)=0$. The $\Pi (Q^2)$
measured on the lattice, in contrast, is unsubtracted, with $\Pi(0)$
determined by the fit. A faithful simulation of the lattice situation may,
however, be obtained by leaving the parameter $\Pi(0)$ appearing in the fit
function (\ref{four}) free and determining its value in the fit~\cite{GMP}.

Vector Meson Dominance-type (VMD-type) fits have been frequently used in
the literature~\cite{VMD}. In its simplest incarnation, VMD corresponds to
taking $K=1$ and $a_0=0$ in Eq.~(\ref{four}) and fixing  $b_1=m_{\rho}^2$.
An extended version, which we will call VMD+, is obtained by making
$a_0$ a free, in general non-zero, fit parameter. It is
important to realize that, despite appearances, VMD-type functions are not
elements of the Pad\'{e} sequence known to converge to the true VP function.
From the point of view of the analytic properties of $\Pi(Q^2)$, whose
discontinuity is produced by a cut starting at $Q^2= - 4 m_{\pi}^2$ rather
than a simple pole, the choice $b_1=m_{\rho}^2$ seems somewhat arbitrary.
Pad\'{e}s converge to the true function because their poles pile up in
such a way that they resemble the cut, and it is hard to imagine how this
could also happen if certain poles are fixed at some predetermined values,
as in VMD. In fact, as we will see in the next section, our model yields
convincing evidence that VMD-type fits are unreliable at the level of
precision desired in the $a_\mu^{HVP}$ problem.

\section{Results}

\begin{table}[t]
\begin{center}
\begin{tabular}{|c|c|c|c|c|}
\hline
&  $\widetilde{a}_{\mu}^{\mathrm{HVP}, Q^2\leq 1} \times 10^7$ & Error$\times 10^7$ & $\chi^2/\mathrm{dof}$ & Pull \\
\hline
VMD & {1.3201} &{ 0.0052} & { 2189/47} & { -} \\
VMD+ & { 1.0658} & { 0.0076} & { 67.4/46} & {18} \\
\hline
$[0,1]$   &   {0.8703}  &  { 0.0095}    &     { 285/46} & { -}\\
$[1,1]$   &  1.116  &   0.022     &      61.4/45 & 4\\
$[1,2]$   & {1.182}  & { 0.043}      &   {55.0/44} & {0.5} \\
$[2,2]$    &  { 1.177} & { 0.058 }    &  { 54.6/43} &{0.5} \\
\hline
\end{tabular}
\caption{Results of different types of fits to a realistic set of lattice data. See text.}
\label{tab1}
\end{center}
\end{table}%

Results of correlated fits to the fake data set,
constructed using the covariance matrix employed to generate that data,
are shown in Table~\ref{tab1}. The first column gives the fit
function form, the second the result obtained for
$\widetilde{a}_{\mu}^{\mathrm{HVP}, Q^2\leq 1}$, using Eq.~(\ref{four})
and the fitted version of that function. The resulting statistical
error on $\widetilde{a}_{\mu}^{\mathrm{HVP}, Q^2\leq 1}$ is shown
in column 3 and the $\chi^2/\mathrm{dof}$ for the fit in column 4. The
last column shows the ``Pull'', a measure of the reliability of the
fit, defined to be the ratio of the true error to the fit error, i.e.,
\begin{equation}\label{pull}
    \mathrm{Pull}=\frac{\mathrm{exact\ value}-\mathrm{fit\ value}}{\mathrm{error}}\ .
\end{equation}
A reliable fit should have an error comparable to the difference between
the exact and fit values, i.e. a Pull $\lesssim 1$. Recall that the
exact result, given in Eq.~(\ref{four}), is
$\widetilde{a}_{\mu}^{\mathrm{HVP}, Q^2\leq 1}= 1.204 \times 10^{-7}$.

From Table~\ref{tab1} we observe that the small error obtained from the VMD
fit is very misleading, though the large  $\chi^2/\mathrm{dof}$ is at
least a warning that the fit is bad and should not be trusted. We have thus
refrained from quoting the corresponding Pull. More worrisome is the result
for VMD+. The fit is apparently good, with a $\chi^2/\mathrm{dof}$ acceptably
close to $1$, but, despite the very small error obtained from the fit,
the value for $\widetilde{a}_{\mu}^{\mathrm{HVP}, Q^2\leq 1}$ is totally
wrong, as emphasized by a very large Pull. The result based on this fit has a
systematic error which the $\chi^2/\mathrm{dof}$ completely fails to expose!

\begin{figure}
\centering
\includegraphics[width=2.6in]{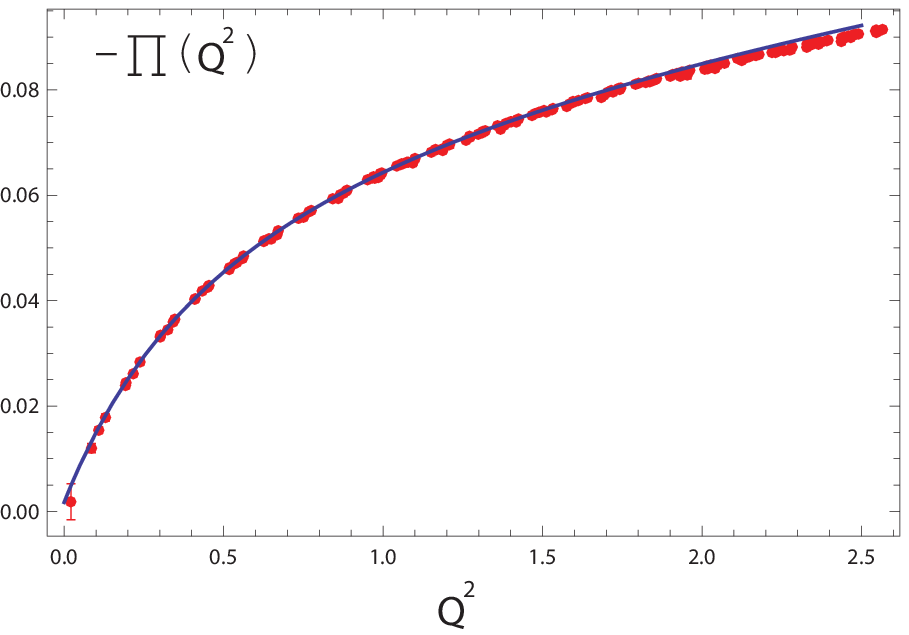}
\hspace{1cm}
\includegraphics[width=2.7in]{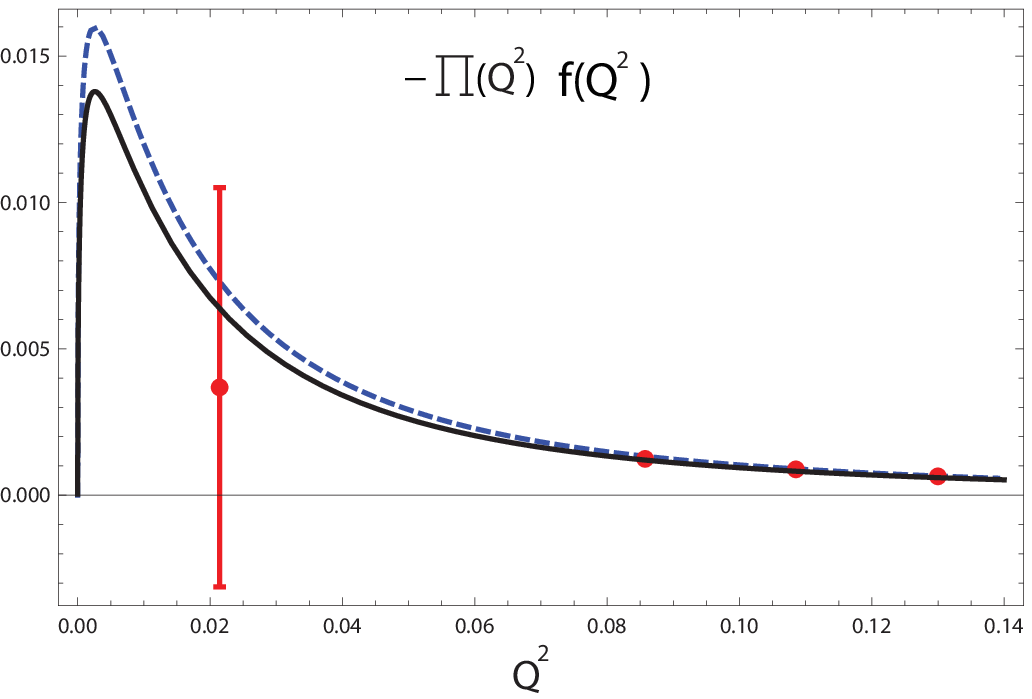}
\caption{Left panel: VMD+ fit to the fake data corresponding to
the model $\Pi(Q^2)$. Right Panel: (Black solid curve) Result of the
fit in the left panel for the integrand of
$\widetilde{a}_{\mu}^{\mathrm{HVP}}$, Eq. (1.1).
(Blue dashed curve) The corresponding exact model result.}
\label{moral}
\end{figure}

The source of the misleading nature of the result for $\widetilde{a}_{\mu}^{\mathrm{HVP}, Q^2\leq 1}$ obtained
from the VMD+ fit can be understood from Fig.~\ref{moral}.
The left panel shows the result of the VMD+ fit for the VP
function $\Pi(Q^2)$. The fit looks very good, as corroborated by the decent
$\chi^2/\mathrm{dof}=67.4/46$ shown in Table~\ref{tab1}. This, however,
obscures the fact (shown in the right panel) that, in the region where
the integrand gets most of its contribution (and where no data exists!)
the fit (depicted by the black solid curve) significantly misrepresents
the exact result (depicted by the blue dashed curve),
leading to the wrong estimate for the integral shown in Table~\ref{tab1},
and the consequent large value, $18$, for the Pull. The moral of this exercise is twofold. First, it is
potentially very misleading to judge the reliability of the value of
$\widetilde{a}_{\mu}^{\mathrm{HVP}, Q^2\leq 1}$ obtained from a given fit
based solely on the agreement of the fit and data versions of the VP
function; a plot of the fit and data versions of the integrand
is safer in this regard. Second, since the source of the problem
is a failure of the fitted version to adequately
represent the curvature of $\Pi (Q^2)$ at those $Q^2$ lying
outside the fit window, and hence an insufficiently accurate extrapolation
from the lowest $Q^2$ data point down to $Q^2$ in the vicinity of the
peak of the integrand, the more precise data one has in the peak
region and/or just above it, the more likely it will be that the
error coming out of the fit will be reliable.

Concerning the Pad\'{e}s, one sees that although the first elements in
the sequence (i.e. the $[0,1]$ and the $[1,1]$ Pad\'{e}s) do not do a
great job, as witnessed by the corresponding $\chi^2/\mathrm{dof}$, by
the time the  $\chi^2/\mathrm{dof}$ is close to $1$ the error
extracted from the fit \emph{is} reliable, giving Pulls $<1$.
This is the case for the $[1,2]$ and $[2,2]$ Pad\'{e}s. Space constraints
preclude displaying the Pad\'{e} analogues of Fig.~\ref{moral}.
The $[1,2]$ Pad\'{e} version is, however, presented in Ref.~\cite{GMP} and
shows a left panel essentially identical to that of Fig.~\ref{moral}.
The right panel, however, is completely different, the solid black
and dashed blue curves now lying almost on top of one another.
The Pad\'{e}s approach thus appears to represent a more reliable
fit strategy. Note, however,
that the final error for $\widetilde{a}_{\mu}^{\mathrm{HVP}, Q^2\leq 1} $
obtained using the $[1,2]$ and $[2,2]$ Pad\'{e}s is $\sim 4-5\%$. Since
other sources of error (chiral and continuum extrapolation, finite
size effects, disconnected diagrams, etc...) have yet to be added,
we conclude that few $\%$ total error estimates for current lattice
determination of $a_{\mu}^{\mathrm{HVP}}$ must be considered unrealistic.

In view of the $\sim 1\%$ error claimed for the determination of
$a_{\mu}^{\mathrm{HVP}}$ based on experimental electroproduction
cross-sections~\cite{exp}, we have also performed the following
exercise. Keeping the data and $\{Q_i^2\}$ set the same, we have
divided the errors by a factor of $\sim$ 100 (i.e., the covariance
matrix by a factor of $10^4$) and then generated (randomly) a new
fake data set. Though clearly not realistic for near-future
lattice determinations, this ``Science Fiction" scenario allows us
to investigate whether such a drastic ``brute-force" error reduction
would allow the lattice to achieve a competitive
($\sim 1\%$) determination of $a_{\mu}^{\mathrm{HVP}}$. The answer
we find is ``barely." The results of the corresponding fits
are shown in Table~\ref{tab2}.

\begin{table}[t]
\begin{center}
\begin{tabular}{|c|c|c|c|c|}
\hline
&  $\widetilde{a}_{\mu}^{\mathrm{HVP}, Q^2\leq 1} \times 10^7$ & Error$\times 10^7$ & $\chi^2/\mathrm{dof}$ & Pull \\
\hline
VMD & {1.31861} &{ 0.00005} & { 2$\times 10^{7}$/47} & { -} \\
VMD+ & { 1.07117} & { 0.00008} & { 7$\times 10^{4}$/46} & - \\
\hline
$[0,1]$   &   {0.87782}  &  { 0.00009}    &     {2$\times 10^{7}$ /46} & { -}\\
$[1,1]$   &  1.0991  &   0.0002     &      5 $\times 10^{4}$/45 & -\\
$[1,2]$   & {1.1623}  & { 0.0004}      &   {1340/44} & {-} \\
$[2,2]$    &  { 1.1862} & { 0.0015 }    &  { 76.4/43} &{12} \\
$[2,3]$    &  { 1.1965} & { 0.0028 }    &  { 42.0/42} &{2} \\
\hline
\end{tabular}
\caption{Results of the same type of fits to a ``Science Fiction" lattice data set, with a 100 times smaller errors.}
\label{tab2}
\end{center}
\vspace*{-4ex}
\end{table}%

One sees in Table~\ref{tab2} that VMD and VMD+ are incapable of producing
a decent fit (i.e., the $\chi^2$ is simply huge). While low-order Pad\'{e}s
have the same problem, the general convergence property comes to the rescue
once sufficiently high orders are reached. This happens for the $[2,3]$
Pad\'{e}. However, even then the Pull is $2$, i.e., the error from
the fit underestimates the true
discrepancy with the exact value ($0.6\%$) by a factor of 2.
The obvious conclusion seems to be that major improvements
in the determination of $a_{\mu}^{\mathrm{HVP}}$ will require a
denser set of data points close to the integrand peak shown in
Fig.~\ref{moral}. This points to the use of
larger lattice volumes and/or twisted boundary conditions.
The recent proposals of Refs.~\cite{new} may also help.

\section{Conclusions}

Current lattice determinations of $a_{\mu}^{\mathrm{HVP}}$ require the use
of a fit function to extrapolate data to the much lower $Q^2$ dominating
the integral representation (\ref{one1}) of $a_{\mu}^{\mathrm{HVP}}$.
Comparisons in a model context, where the exact value is known, show that
VMD-type fits are not reliable, failing to achieve an accuracy of even a
few $\%$. Pad\'{e} approximants of sufficiently high order
represent a more promising approach.

We have shown that, as a consequence of the non-trivial curvature
of $\Pi (Q^2)$ in the low-$Q^2$ region, the $\chi^2/\mathrm{dof}$
of a fit cannot, in general, be used to assess the reliability of
the associated result for $a_{\mu}^{\mathrm{HVP}}$ unless very good
data is available in and/or near the region of the peak of the integrand in
Eq.~(\ref{one1}). Plots of the fit to $\Pi(Q^2)$ may thus be very
misleading, and alternate forms showing the data and fit versions
of the integrand in Eq.~(\ref{one1}) serve better to reveal
the quality of those aspects of the fit to the data most relevant to
a reliable determination of $a_{\mu}^{\mathrm{HVP}}$.

In lattice calculations, where the underlying exact $\Pi (Q^2)$ is not
known {\it a priori}, our results show the importance of a quantitative
assessment of the systematic error associated with the use of particular
fit functions and methods. We encourage all future lattice calculations
to include such assessments, which can be obtained by using the $Q^2$
values and covariance matrix of the lattice data entering the fits to
generate a set of fake data with the model for $\Pi(Q^2)$ presented here.
While achieving a $\lesssim 1\%$ lattice determination of
$a_{\mu}^{\mathrm{HVP}}$ won't be a rose garden, it is very important to try.

\end{document}